\documentstyle[prl,preprint,aps]{revtex}
\begin{document}
\draft
\title{Quantized thermal conductance of dielectric quantum wires}
\author{Luis G. C. Rego and George Kirczenow}
\address{Department of Physics, Simon Fraser University,
Burnaby, B.C., Canada  V5A 1S6}
\date{Received 10 December 1997}
\maketitle
\begin{abstract}

Using the Landauer formulation of transport theory, we predict that
dielectric quantum wires
should exhibit quantized thermal conductance  at low temperatures
in a ballistic phonon regime. The quantum of thermal
conductance is universal, independent of the characteristics of
the material, and equal to
$\pi^2 k_B^2 T/3h$ where $k_B$ is the Boltzmann constant,
$h$ is Planck's constant and $T$ is the temperature.
Quantized thermal conductance
should be experimentally observable
in suspended nanostructures adiabatically
coupled to reservoirs, devices that can be realized at the
present time.

\end{abstract}
\pacs{PACS: 85.30.Vw, 73.23.Ad,
63.22.+m}
%\begin{multicols}{2}

During the last two decades, the physics of electron transport in one
dimension has attracted a great deal of attention. Some remarkable
associated phenomena
have been the quantum Hall effect discovered by von Klitzing, Dorda and
Pepper\cite{Kli} and the quantized conductance of ballistic point
contacts discovered by van Wees {\em et al}.\cite{Wees88} and Wharam
{\em et al}.\cite{Wha}
Here the signature of one-dimensional conduction has been the quantization
of the two-terminal and Hall electrical conductances in multiples
of the fundamental quantum $e^2/h$, as has been understood within the
framework of Landauer theory,\cite{Lan57} and of
B\"{u}ttiker-Landauer theory,\cite{BuLan} respectively.

One-dimensional phonon transport should also be possible. However, despite
the long standing theoretical interest in this topic that goes back to
the 1920's,\cite{Peierls}
the question whether the phonon thermal conductance
should be quantized in one dimension has to our knowledge not
been addressed either theoretically or experimentally. Recent advances
in nanotechnology have made experimental investigation of this question 
feasible; we note especially the detection of nanowire phonon subbands 
by Seyler and Wybourne\cite{Wyb92} and the measurement of the 
thermal conductance of a suspended nanostructure by Tighe, Worlock and 
Roukes\cite{Rou97}.

The purpose of this Letter is to demonstrate theoretically that in
a low temperature
regime dominated by ballistic massless phonon modes
the phonon thermal conductance
of a one dimensional quantum wire is quantized, the fundamental
quantum of thermal conductance being $\pi^2 k_B^2 T/3h$, where
$k_B$ is the Boltzmann constant,
$h$ is Planck's constant and $T$ is the temperature. We also establish
the conditions that should be met for the experimental observation
of this novel phenomenon.

Our starting point is the Landauer energy flux
\begin{eqnarray}
\dot{Q} = \sum_{\alpha} \int_{0}^{\infty}
    \frac{dk}{2\pi}\ \hbar \omega_{\alpha}(k) \ v_{\alpha}(k)
    \left( \eta_R - \eta_L \right)
    \zeta_{\alpha}(k) \
\label{q1}
\end{eqnarray}
carried by a quantum wire
connecting two reservoirs labeled R and L. Here
$\omega_{\alpha}(k)$ and $v_{\alpha}(k)$ are the frequency and
velocity of normal mode ${\alpha}$ of the quantum wire
with wave-vector $k$, $\zeta_{\alpha}(k)$ is the phonon
transmission
probability through the wire and
$\eta _i (\omega) = 1/(e^{h\omega/k_BT_i}-1)$
represents the
thermal distribution of phonons in the reservoirs, assumed to be a Planck
distribution at temperature $T_i$.
The cross sectional area of the wire is assumed to be of the order
of hundreds of $nm^2$, so that the lateral confinement produces finite gaps
in the dispersion relation of the phonon frequencies. Eq. (\ref{q1})
transforms to
\begin{eqnarray}
\dot{Q} = \frac{1}{2\pi} \sum_{\alpha} \int_{\omega_{\alpha}(0)}^{\infty}
          d\omega \ \hbar \omega \left( \eta_R(\omega) - \eta_L(\omega)
\right)
          \zeta_{\alpha}(\omega) \
\label{q3}
\end{eqnarray}
since the phonon velocity
$v_{\alpha}(k)=\partial \omega_{\alpha}/\partial k$ is canceled by the
1D density of states
$g(\omega_{\alpha})= \partial k/\partial \omega_{\alpha}$.

The reservoir-to-reservoir thermal conductance of the wire
is $\kappa = \dot{Q}/\Delta T$,
where $\Delta T=T_R - T_L$ is
the temperature difference between the reservoirs. It follows that
%\end{multicols}
\begin{eqnarray}
\kappa =  \frac{1}{2\pi} \left\{ \sum_{\alpha}^{N_{\alpha}} \int_0^{\infty}
           d\omega \ \hbar \omega
           \left( \frac{\eta_R(\omega) - \eta_L(\omega)}{\Delta T} \right)
           \zeta_{\alpha}(\omega)
     \ +   \ \sum_{{\alpha}'}^{N_{{\alpha}'}}
\int_{\omega_{\alpha'}(0)}^{\infty}
           d\omega\ \hbar \omega
           \left( \frac{\eta_R(\omega) - \eta_L(\omega)}{\Delta T} \right)
           \zeta_{{\alpha}'}(\omega) \right\} \ .
\label{q5}
\end{eqnarray}
%\begin{multicols}{2}
We have separated Eq.(\ref{q5}) in two parts, so that the first term
represents the conductance of the massless modes
with $\omega_{\alpha}(0)=0$ and the second is the contribution to the
thermal conductance due to the higher energy modes,
which have a finite cut-off
frequency $\omega_{\alpha'}(0) \not = 0$.

For the moment let us assume perfectly adiabatic contact between the
thermal reservoirs and the ballistic quantum wire,
so that $\zeta(\omega)=1$.
For this idealized case integration of Eq.(\ref{q5}) yields
%\end{multicols}
\begin{eqnarray}
\kappa = \frac{k_B^2 \pi^2}{3h}
           \left(\frac{T_R+T_L}{2}\right) N_{\alpha}
       + \frac{k_B^2}{h} \ \sum_{{\alpha}'}^{N_{{\alpha}'}} \left\{
       \frac{\pi^2}{3} \left(\frac{T_R+T_L}{2}\right)
       + \frac{1}{\Delta T} \left[ T^2_R dilog(e^{x_{R{\alpha}'}})
       - T^2_L dilog(e^{x_{L{\alpha}'}}) \right]  \right\} \ ,
\label{q6}
\end{eqnarray}
%\begin{multicols}{2}
where $x_{(R,L){\alpha}'}=\hbar \omega_{{\alpha}'} (0) /k_B T_{(R,L)}$,
for the modes with finite cut-off frequencies.

Equation (\ref{q6}) predicts a remarkable behavior for the thermal transport
properties of 1D phonon systems:
Each massless mode presents a {\em universal thermal
conductance} equal to the product of the averaged reservoir temperature
and the universal constant $k_B^2 \pi^2/3h$. The higher energy modes,
on the other hand, show a dependence on the intrinsic properties of the
material and on the geometrical parameters of the sample through the
cut-off frequencies $\omega_{{\alpha}'}(0)$. However their contribution
to $\kappa$ is exponentially small at low temperatures. In the limit
$\Delta T \rightarrow 0$ we find
\begin{eqnarray}
\kappa = \frac{k_B^2 \pi^2}{3h} T N_{\alpha}
       + \frac{k_B^2}{h} T \ \sum_{{\alpha}'}^{N_{{\alpha}'}}
           \left\{ \frac{\pi^2}{3} + f(x_0)
           + \frac{x_0^2 e^{x_0}}{e^{x_0}-1} \right\} \
\label{q7}
\end{eqnarray}
with $f(x)=2 dilog(e^{x})$ and
$x_0 = \hbar \omega_{{\alpha}'} (0) / k_B T$.

Expressions (\ref{q6}) and (\ref{q7}) represent an idealized case in
which the transmission of phonons from one reservoir to the
other happens without reflections. A more realistic model has to incorporate
the effects of the reflections caused by the contacts between the reservoirs
and the 1D wire. The remainder of this paper is devoted to the study of
realistic physical systems in which the universal thermal conductance of
phonons could be observed. The systems we consider are similar to the
experimental device of Tighe, Worlock and Roukes\cite{Rou97} where a quasi-1D
quantum wire connects quasi-2D reservoirs.

We  begin our analysis by examining the dispersion relations
of low energy phonons in a quasi-2D system. At low
temperatures
of the order of a few degrees Kelvin, the dominant phonon wavelength is much
larger than the lattice parameter and the model of an elastic continuum
can be used. We assume an isotropic crystal, which can represent
Si or GaAs at low temperatures to a good approximation.
The optical phonons are not considered because of their high energy. In this
model Rayleigh-Lamb modes\cite{Auld}
describe the quasi-2D acoustic phonons. Figure \ref{fig1} shows the frequency
spectra of the lowest energy symmetric (solid lines) and anti-symmetric
(dashed lines) modes of a
quasi-2D system of $50nm$ thickness. The parameters used are those of
GaAs. The symmetries of the modes refer
to atomic displacements in the $z$ direction,
perpendicular to the quasi-2D system.
The energy scale is degrees Kelvin, so that one may estimate
the temperature at which the modes with finite cut-off frequency begin to
contribute to thermal transport. These cut-off frequencies increase as the
sample thickness is reduced.
Together with the 2D modes of the reservoirs we have plotted
in Fig. \ref{fig1} the lowest energy branches of the longitudinal
(circles), transverse
(squares), shear (diamonds) and torsional modes (triangles)
of a quasi-1D quantum wire with a
square cross section of $50nm\times 50nm$.
The fundamental torsional mode is given by the simple beam theory,
which is well justified by the fact that 
numerical calculations\cite{Nigro} show that its phase velocity is
practically constant for all frequencies.
The transverse and shear modes are coupled by the Timoshenko
equation\cite{Graff}.
These quasi-1D modes agree very well at low
energies with the 2D modes of the reservoirs. Therefore they
should propagate readily from the reservoirs to
the wire. Since expressions (\ref{q6}) and
(\ref{q7}) do not depend on the phonon velocity, but only on the cut-off
frequencies, the modes shown in Fig. \ref{fig1} represent
the system very well up to temperatures of the order of 1K.

Nevertheless, there still remains the question of
how reflections would change the
results given in  Eq. (\ref{q6}) and (\ref{q7}).
We will discuss this here in detail for the case of
the longitudinal modes. For a
quasi-1D wire whose cross sectional dimensions are much smaller than its
length, the transmitted modes are well described by the normal modes of a
long beam. Let us suppose that the cross
section of the beam varies along
its length, assumed to be in the $x$ direction. The equation of motion
of a longitudinal plane wave traveling along the $x$
direction is given by\cite{Graff}
\begin{eqnarray}
\frac{\partial^2 u}{\partial x^2} +
\frac{1}{A(x)}\frac{\partial A(x)}{\partial x}\frac{\partial u}{\partial x}
= \frac{1}{v_l^2}\frac{\partial^2 u}{\partial t^2} \ .
\label{q8}
\end{eqnarray}
$A(x)$ is the cross sectional
area of the beam and $v_l$ is the velocity of the longitudinal
mode $v_l=\sqrt{Y/\rho}$, $Y$ is
Young's modulus and $\rho$ the density.
To solve Eq.(\ref{q8}) we must specify $A(x)$, so that
the transmission coefficients $\zeta(w)$ are determined by the shape of
the contacts between the reservoirs and the 1D wire.
We consider two contact geometries: conical, for which
$A(x)=A_0 tan(\theta)(x + x')$, and catenoidal where
$A(x) = A_0 cosh^2(x/\lambda)$.
These are illustrated in Fig. \ref{fig2} by schemes (A) and (B),
respectively.
$\theta$ is the angle of
flare of the conical contact and $\lambda$ is the characteristic length of the
catenoid.
The broad regions at both extremes of the
structures are 40 times wider than the 1D channel ($50\ nm$ wide).
The thickness (in the direction perpendicular to the plane of the figure, $z$)
is constant  along the whole structure and chosen to be $50nm$.
Considering initially the case of conical contacts, two distinct
wave equations result for the straight (I) and conical (II) parts of the
structure
\begin{eqnarray}
\frac{\partial^2 u_I}{\partial x^2}
&=& \frac{1}{v_l^2}\frac{\partial^2 u_I}{\partial t^2} \\
\frac{\partial^2 u_{II}}{\partial x^2} +
\frac{1}{x+x'}\frac{\partial u_{II}}{\partial x}
&=& \frac{1}{v_l^2}\frac{\partial^2 u_{II}}{\partial t^2}  \  ,
\label{q9}
\end{eqnarray}
with solutions
\begin{eqnarray}
u_I(x,t) &=& Be^{i(\pm kx-wt)} \\
u_{II}(x,t) &=& C \left[ J_0(k(x+x')) \pm iN_0(k(x+x')) \right] e^{-iwt} \ .
\label{q10}
\end{eqnarray}

These solutions are matched at the interfaces between the conical and straight
regions, where the continuity of the particle velocities $\dot{u}(x,t)$ and
stresses $\sigma = A(x)Y\frac{\partial u(x,t)}{\partial x}$ are
required. These boundary
conditions lead us to a system of coupled equations that is solved for
the transmission coefficients of plane waves traveling along
the structure. Fig.\ref{fig2} shows the transmission coefficient $\zeta(k)$
as a function of the longitudinal wave-vector $k$ of a structure with
conical contacts in
which $\theta=\pi /6$ (solid line).
The peaks are resonances associated with reflections at the ends of
the conical regions and of the 1D wire.
For $k=0$ the transmission coefficient is equal to one. The
overall behavior of $\zeta(k)$ is determined by the conical shape of
the contact, that guarantees a finite transmission for all frequencies, but
is inefficient for a wide range of frequencies.

The catenoidal contacts present a better response for lower values of $k$.
In this case the equations that define the transmitted wave are
\begin{eqnarray}
\frac{\partial^2 u_I}{\partial x^2}
&=& \frac{1}{v_l^2}\frac{\partial^2 u_I}{\partial t^2} \\
\frac{\partial^2 u_{II}}{\partial x^2} +
\frac{2 tanh(x/\lambda)}{\lambda}\frac{\partial u_{II}}{\partial x}
&=& \frac{1}{v_l^2}\frac{\partial^2 u_{II}}{\partial t^2} \ ,
\label{q11}
\end{eqnarray}
with solutions
\begin{eqnarray}
u_I(x,t) &=& Be^{i(\pm kx-wt)} \\
u_{II}(x,t) &=& {C e^{i(\pm kx-wt)}}({cosh(x/\lambda)})^{-1}  \ .
\label{q12}
\end{eqnarray}
The transmission coefficients obtained in this case are illustrated by the
dashed curve in Fig. \ref{fig2}. Substitution of the function $u_{II}$
in Eq.(\ref{q11}) results in the frequency spectrum
$\omega^2=v_l^2(k^2+(1/\lambda)^2)$
for a wave traveling along the catenoidal contact,
which has a cut-off at frequency $\omega_0=v_l/\lambda$.
For $\omega<\omega_0$ the wave is evanescent,
but a resonance guarantees unitary
transmission for $k$=0. However the cut-off frequency becomes
smaller as the
parameter
$\lambda$ that characterizes the length of the catenoid
increases.
The  efficiency of transmission of longitudinal
waves in this system is also limited by  reflections that happen
on passing between the straight region and the curved one.
These reflections can be avoided for a catenoidal contact that is not
laterally limited.
This type of contact is exemplified by the structure (C) in Fig. \ref{fig2},
the {\it infinite catenoidal} contact. The transmission $\zeta(k)$ for
this case is 1 for all wave vectors $k$, but the cut-off at
$\omega_0=v_l/\lambda$ still affects the overall transmission.

With a knowledge of $\zeta(\omega)$ it is possible to obtain realistic
results for the 1D thermal conductance. Initially we analyze the case of
ideal transmission for all modes, which is a
very good approximation for the
large $k$ acoustic waves that dominate the thermal transport between 100mK
and 1K.
Six distinct modes contribute to the thermal
transport through the wire below 1K: a
longitudinal mode, 2 transverse modes, a torsional mode and
2 shear modes.
Among these only the shear modes present  a
cut-off ($\hbar \omega_0=1.62 K$) that is the result of lateral confinement
(see Fig. \ref{fig1}).
It was verified that the higher branches of these modes
do not contribute significantly
to the thermal transport at temperatures lower than 1K.
We considered the limit $\Delta T \rightarrow 0$ of Eq.(\ref{q7}) and
calculated the ideal thermal conductance divided by temperature of the
quantum wire. The result is shown in Fig. \ref{fig3} (refer to the left scale).
For the ideal contact, at low temperatures, this yields
the quantized thermal
conductance $k_B^2\pi^2/3h$ times the number of massless
modes ($N_{\alpha}$=4). As the
temperature approaches 1K the shear modes begin to contribute to the process
(increasing the thermal conductance) and the plateau that is the signature
of quantized thermal conductance terminates.
Now considering the transmission coefficient of the longitudinal mode, we
have plotted its contribution to the thermal conductance
for various contact shapes (refer to the right scale of Fig. \ref{fig3}).
In this case the plateau is
modified at the low temperature side by reflections at the contacts
and, to a smaller extent, by the cut-off of the frequency spectrum
of the catenoidal contacts. Nonetheless for the catenoidal
contacts a very distinctive plateau characteristic
of quantized thermal conductance is clearly visible over a wide range of
temperatures. Therefore based on the behavior of the longitudinal mode,
a signature of the quantized thermal conductance of dielectric quantum
wires can be expected between 30mK and 300mK.
We note that the principle of
adiabatic matching that underlies our discussion of the longitudinal
mode is very general and applies to the other massless modes as
well. Therefore these should exhibit qualitatively similar behavior,
including universal quantized thermal conductance over a similar range
of temperatures. Since adiabatic wave propagation does 
not require an isotropic medium, our predictions apply even
if there is phonon focusing due to crystal anisotropy.  
Similar results also apply to wires with cylindrical symmetry which 
admit analytic solutions for all of the modes,
but will be more difficult to fabricate.

In this study we did not consider the effects of surface
roughness\cite{Rou97}
and other defects that
may limit the transmission of phonons through the quantum
wire.
However these are technological limitations that should be overcome
with improved control of the growth process. 
They should also be of less concern at lower 
temperatures.\cite{Klitsner}
Furthermore, studies of
1D electron transport have shown that quantization
plateaus can exist in the presence of defect scattering under
certain conditions.
\cite{Nixon}.

In summary, we predict the existence in 1D systems of a universal
quantum of thermal
conductance due to phonons that is equal to $k_B^2\pi^2/3h$
times the temperature, for the lowest energy modes. The conductance of
the higher energy modes is influenced by the geometrical and intrinsic
parameters of the system through the gaps in the frequency spectrum.
Realistic structures were investigated and it was found that contacts
with catenoidal shape should be those that better represent the ideal
case in experiments. Finally, we note that the Wiedemann-Franz law 
applies in 1D\cite{WF1},\cite{WF2} and has been tested 
experimentally for 
ballistic point contacts.\cite{WF3} It predicts\cite{WF2} 
that the 1D thermal 
conductance of ballistic {\em electrons} should also be quantized 
in multiples of $k_B^2\pi^2 T/3h$. Thus we arrive at the surprising 
conclusion that the low temperature 1D thermal conductance of 
ballistic phonons and electrons is described by the same universal 
quantum even though bosons and fermions obey different statistics.

We thank Michael Roukes for stimulating discussions. This work was
supported by NSERC of Canada.

%\end{multicols}
%%%%%%%%%%%%%%%%%%%   FIGURE  CAPTIONS  %%%%%%%%%%%%%%%%%%%%%%%%%%%%%%%

\begin{figure}
\caption{ Lowest energy acoustic modes of a quasi-2D system of
thickness $50nm$ and of a long wire of square cross section
$50nm\times50nm$. Thick solid (dashed) lines represent the symmetric
(anti-symmetric) modes of the quasi-2D system. Thin lines with circles,
squares, diamonds and triangles represent the longitudinal, transverse, shear
and torsional modes of the wire, respectively.
The elastic parameters are for GaAs.}
\label{fig1}
\end{figure}

\begin{figure}
\caption{Transmission coefficients vs. longitudinal wave-vector
$k$. A) conical contact, $\theta=\pi/6$; the quasi-1D wire at the
center of the structure has cross-section $50nm\times 50nm$ and length
$1\mu$. B) finite catenoidal contact of characteristic length
$\lambda = 0.86 \mu$.
C) {\it infinite catenoidal} contact of $\lambda = 0.86 \mu$.
The straight portions at the ends of structures (A) and (B) have width
$2\mu$.}
\label{fig2}
\end{figure}

\begin{figure}
\caption{ Left scale: Thermal conductance of a quantum wire with ideal contacts
divided by temperature. Right scale: contribution to
thermal conductance due to the longitudinal mode for various contact
shapes: {\em infinite catenoid} for $\lambda =4.6 \mu$ (solid), finite
catenoid for $\lambda = 4.6 \mu$ (dot-dashed), catenoid with $\lambda =
0.86 \mu$
(long-dashed) and conic for $\theta = \pi/6$ (solid with circles).
$\kappa^l_{ideal}=k^2_B\pi^2T/3h$.}
\label{fig3}
\end{figure}

%%%%%%%%%%%%%%%%%%%%%%%%%%%%%%%%%%%%%%%%%%%%%%%%%%%%%%%%%%%%%%%%%%%%%%%%%

%%%%%%%%%%%%%%%%%%%%%%%%%%%%%%%%%%%%%%%%%%%%%%%%%%%%%%%%%%%%%%%%%%%%%%%%%%%%%

\end{document}